# A study of trends in the effects of TV ratings and social media (Twitter) -Case study 1-

Takuya Ueoka [1]
Gunma University,Faculty of Social and Information Studies
4-2Aramaki,,Maebashi-city,Gunma 371-0044
s1810002@gunma-u.ac.jp

Yasuko Kawahata[1]
Gunma University,Faculty of Social and Information Studies
4-2Aramaki,,Maebashi-city,Gunma 371-0044
kawahata@gunma-u.ac.jp

Akira Ishii[2]
Tottori University,Department of Applied Mathematics and Physics
4-101Koyama,Tottori-city,Tottori 680-8552
ishii@damp.tottori-u.ac.jp

*Abstract*—The Japanese TV program "Drama A" is a drama broadcast from October to December 2016. The audience rating was sluggish, but this drama marked a high audience rating in 2016. Since it was popular from the middle, and it was speculated that there was a part related to social media in the popularity, we considered existing research methods as a case study. In this paper, we used a mathematical model of the hit phenomenon to examine the impact of audience assessment from social media from a sociophysical perspective. We got the same consideration as the audience rating per minute of video research. This paper is IEEE BIGDATA2018's Revised paper(Consideration on TV audience rating and influence of social media).

## I. INTRODUCTION

When quantitatively judging the fun of TV, we use audience rating. With an audience rating of every minute, it is possible to grasp the movement of the viewer and the peak of the number of viewers. However, due to the development and spread of information transmission equipment, the influence of social media and others has come to be considered. Therefore, in this research, we will examine what kind of relation is existed by dealing with mathematical model of hit phenomenon together with viewer rating.

## II. EQUATION

It is a hit model of mathematical phenomena used in this study.

$$\frac{dI(t)}{dt} = \sum_\xi c_\xi A_\xi(t) + DI(t) + PI^2(t). \quad (1)$$

The left side represents time change of interest of people of a certain topic, and I (t) represents interests of people. The first term on the right side expresses the influence from the media, and in this research the medium thinks as the effect of TV and Web News. The second term represents D of the direct communication of conversation effect, and the third term represents indirect communication P of rumor effect. The coefficients of c, D, P in the mathematical model of this hit phenomenon are called parameters.

## III. ANALYSIS METHOD

Analysis was carried out according to the following procedure.

1. Data collection and Analytics(from Prof.Ishii)

2. Analyze the collected data with a mathematical model (Python)

3. Consideration from calculation result and transition of audience ratings.

Regarding the analysis period, calculations were carried out by dividing it into two types, from the first day of broadcasting of the drama to the last week and from the broadcasting day of the drama until the day before the next broadcasting day.

## IV. ANALYSIS RESULT

Figure 1 shows the audience rating per minute during drama broadcast. Changes appeared in the graph of audience ratings especially from the fourth talk to the sixth talk.

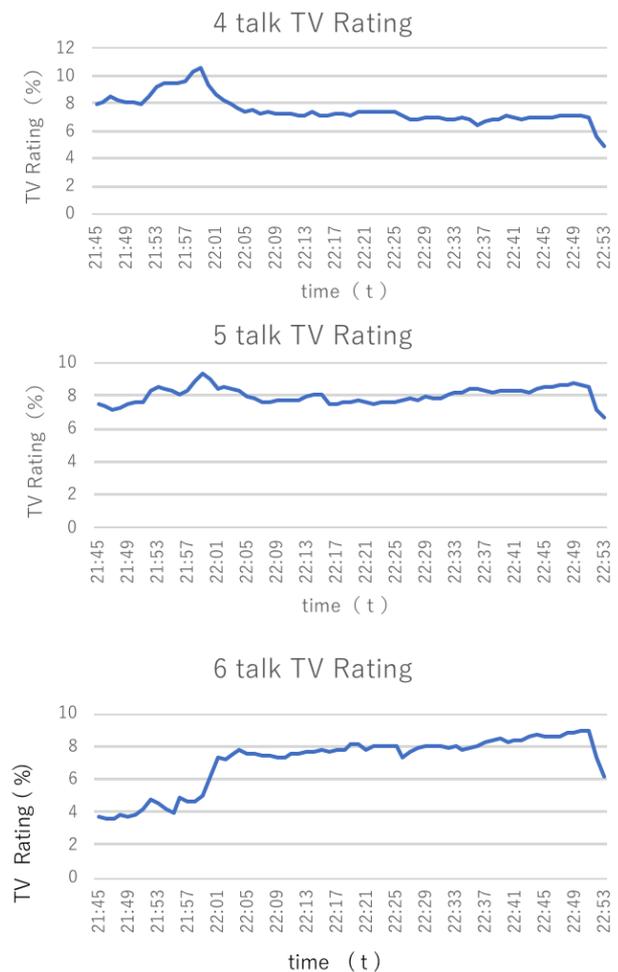

Fig.1 Every minute audience rating of the drama "Drama A". Vertical axis audience rating, horizontal axis time.

From Figure 1, the viewer rating is decreasing when the program starts in the fourth episode of broadcasting. From the fifth episode, the change in the audience rating becomes gentle, and from the sixth talk it turns out that the audience rating is higher than the previous program. From the viewer's rating graph, it seems that there is a popularity factor of the program on the 4-6 talk broadcast.

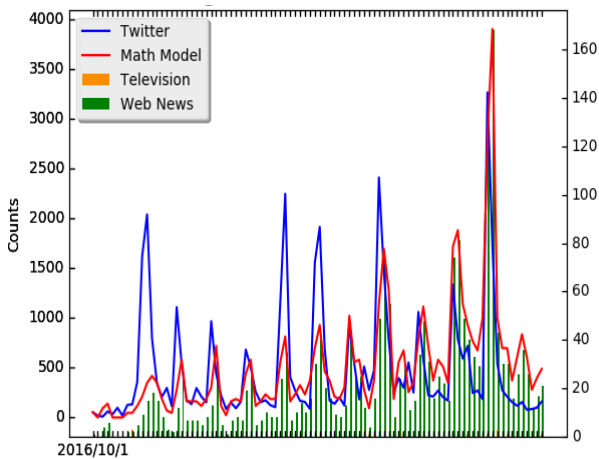

Fig.2 Calculation result using mathematical model of hit phenomenon of drama "Drama A". The red line is reproduced by mathematical model, the blue line and the green bar graph are the number of Twitter and Web news respectively.

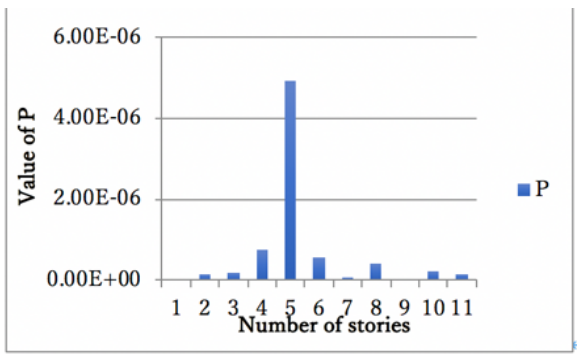

Fig.3 Calculation result using mathematical model of hit phenomenon of "Actor A". The parameter P. The vertical axis shows the value of P and the horizontal axis shows the number of stories

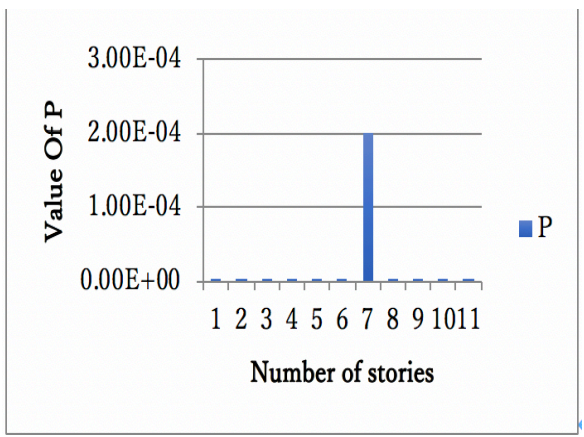

Fig.4 Calculation result using mathematical model of hit phenomenon of "Content". The parameter P. The vertical axis shows the value of P and the horizontal axis shows the number of stories

## V. CONSIDERATION

As shown in Fig. 2, the portion that is the vertex of each graph is the broadcast day of escape shame, and the red line is the result of the mathematical model of hit phenomenon. When paying attention to it, it is getting a steadily rising graph every time we repeat the broadcasting times. For that reason you can see that the popularity of the drama is gradually spreading.

In order to elucidate this popularity, we consider that there are popular factors in the 4th to 6th talks from the viewer rating data in Figure 1 and pay attention to the calculation result by the mathematical model of hit phenomenon. Then, Figures 3 and 4 show the values which are particularly high near the 4th to 6th episodes of broadcasting. First, it seems that there was a factor that increased the expectation degree for something Actor A in the fifth episode as the value of indirect communication P in Fig. 3 " Actor A " was the highest in the fifth episode of broadcasting. In this 5th week 's broadcast, We think that the viewer' s expectation has increased due to the scene where Actor A hugs for the first time and the next time it is a honeymoon.

Next, referring to FIG. 4, P is higher in the seventh talk, not the 4-6 talk. In other words, after the popularity of the program has risen as the viewer rating rises, the value of P increases. It shows that popularity of drama has increased as "popular love dance" became popular after the awareness of the program increased. This "Content" is a dance of the theme song of the program, celebrities and civilians unrelated to the program dance this love dance and uploading to the SNS so that popularity has grown at a stroke from the 7th week It is thought that it is not

## VI. CONCLUSION

The drama "Drama A", the content of the program and "Actor A" were evaluated and the popularity increased. In that love dance became the popular trigger of the final program. Based on the above analysis, it seems that there is a link between SNS and TV viewer ratings, so it is important to increase the effect of SNS in order to improve the audience rating of TV programs.